\documentclass[aps,prd,twocolumn,preprintnumbers,superscriptaddress,nofootinbib,floatfix]{revtex4-1}

\usepackage{graphicx,epstopdf}
\usepackage{amsmath,amssymb,graphicx}
\usepackage[usenames, dvipsnames]{color}
\usepackage{times}
\usepackage{array}
\usepackage{latexsym}
\usepackage{floatrow}
\usepackage{longtable}
\usepackage[utf8, latin9]{inputenc}
\usepackage{hyperref}
\definecolor{steelblue}{RGB}{25,25,112}
\definecolor{DarkGreen}{rgb}{0.0,0.5,0.0}
\hypersetup{colorlinks,linkcolor={blue},citecolor={blue},urlcolor={steelblue}}
\newcommand{\orcid}[1]{\href{https://orcid.org/#1}{\includegraphics[width=8pt]{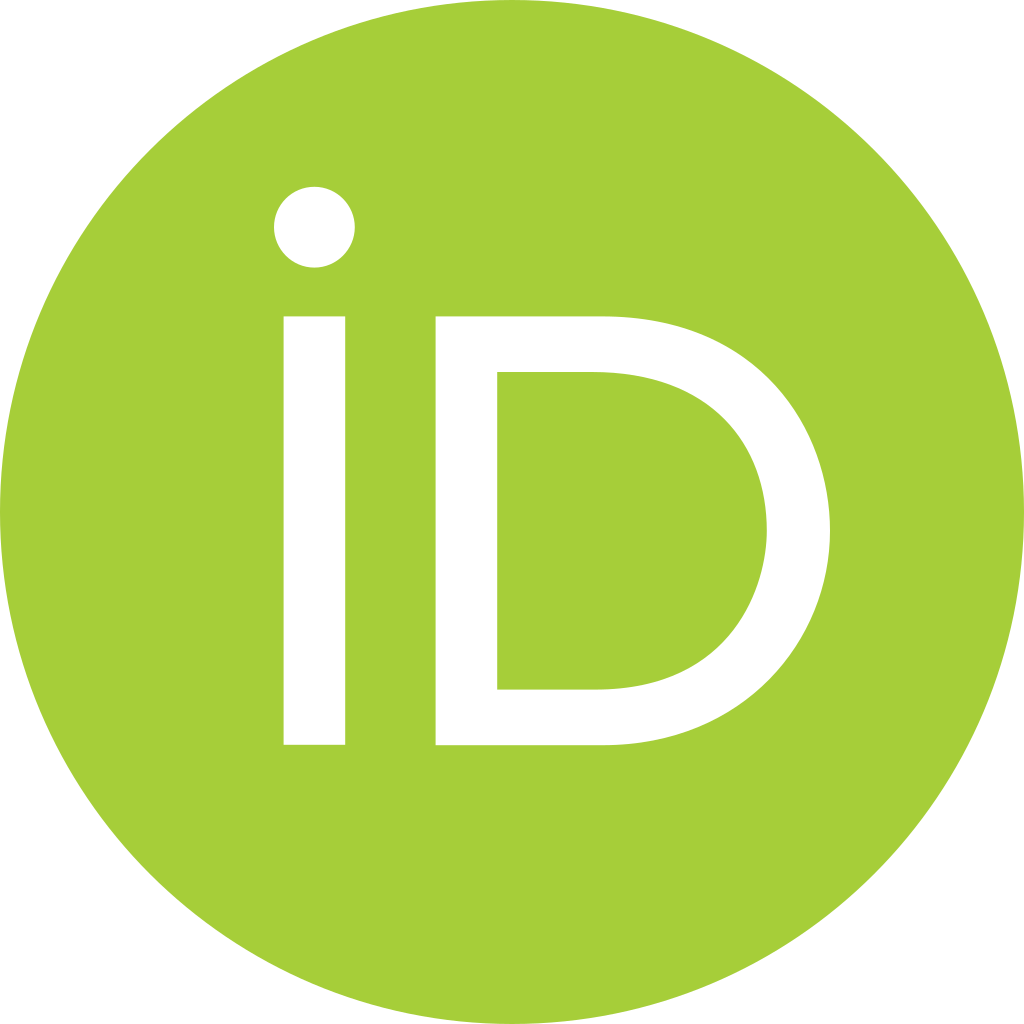}}}

\newcolumntype{L}[1]{>{\raggedright\let\newline\\\arraybackslash\hspace{0pt}}m{#1}}
\newcolumntype{C}[1]{>{\centering\let\newline\\\arraybackslash\hspace{0pt}}m{#1}}
\newcolumntype{R}[1]{>{\raggedleft\let\newline\\\arraybackslash\hspace{0pt}}m{#1}}

\begin{document}
\title{Dark stars powered by self-interacting dark matter}

\newcommand{\LCTP}{\affiliation{
    Leinweber Center for Theoretical Physics, 
    Department of Physics, 
    University of Michigan, 
    Ann Arbor, Michigan 48109, 
    USA}}
\newcommand{\AUSTINAFF}{\affiliation{
    Department of Physics,
	University of Texas, 
	Austin, Texas 78712, 
	USA}}
\newcommand{\STANFORD}{\affiliation{
    Stanford Institute for Theoretical Physics, 
    Stanford University, 
    Stanford, California 94305, 
    USA}}
\newcommand{\OKC}{\affiliation{
    The Oskar Klein Centre for Cosmoparticle Physics, 
    Department of Physics, 
    Stockholm University, \\
    AlbaNova, 
    Roslagstullsbacken 21, 
    10691 Stockholm, 
    Sweden}}
\newcommand{\NORDITA}{\affiliation{
    Nordita, 
    KTH Royal Institute of Technology and Stockholm University, 
    Roslagstullsbacken 23, 
    10691 Stockholm, 
    Sweden}}
\newcommand{\UCR}{\affiliation{
    Department of Physics \& Astronomy, 
    University of California, 
    Riverside, California 92521, 
    USA}}
\newcommand{\TDLI}{\affiliation{Tsung-Dao Lee Institute (TDLI),
    520 Shengrong Road, 201210 Shanghai, People's Republic of China}}
\newcommand{\SJTU}{\affiliation{School of Physics and Astronomy, Shanghai Jiao Tong University,
    800 Dongchuan Road, 200240 Shanghai, People's Republic of China}}

\author{Youjia Wu \orcid{0000-0002-7399-1306}}\email[Electronic address: ]{youjiawu@umich.edu}\LCTP
\author{Sebastian Baum \orcid{0000-0001-6792-9381}}\email[Electronic address: ]{sbaum@stanford.edu}\STANFORD
\author{Katherine Freese}\email[Electronic address: ]{ktfreese@utexas.edu}\AUSTINAFF \OKC \NORDITA
\author{Luca Visinelli \orcid{0000-0001-7958-8940}}\email[Electronic address: ]{luca.visinelli@sjtu.edu.cn} \TDLI\SJTU
\author{Hai-Bo Yu \orcid{0000-0002-8421-8597}}\email[Electronic address: ]{haiboyu@ucr.edu} \UCR

\date{\today}
\preprint{UTTG 12-2021; LCTP-21-18; NORDITA 2021-068}

\begin{abstract}
Dark matter annihilation might power the first luminous stars in the Universe. These types of stars, known as dark stars, could form in $(10^6\textup{--}10^8)\,M_\odot$ protohalos at redshifts $z \sim 20$, and they could be much more luminous and larger in size than ordinary stars powered by nuclear fusion. We investigate the formation of dark stars in the self-interacting dark matter (SIDM) scenario. We present a concrete particle physics model of SIDM that can simultaneously give rise to the observed dark matter density, satisfy constraints from astrophysical and terrestrial searches, and address the various small-scale problems of collisionless dark matter via the self-interactions. In this model, the power from dark matter annihilation is deposited in the baryonic gas in environments where dark stars could form. We further study the evolution of SIDM density profiles in the protohalos at $z \sim 20$. As the baryon cloud collapses due to the various cooling processes, the deepening gravitational potential can speed up gravothermal evolution of the SIDM halo, yielding sufficiently high dark matter densities for dark stars to form. We find that SIDM-powered dark stars can have similar properties, such as their luminosity and size, as dark stars predicted in collisionless dark matter models.
\end{abstract}

\maketitle

\section{Introduction}
\label{sec:intro}

The first stars are thought to have formed inside $(10^6\textup{--}10^8)\,M_\odot$ dark matter halos at redshifts $z \sim 20$ when the Universe was about $200$ million years old~\cite{Abel:2001pr, Gao:2006ug,Wise:2007bf}. In the standard picture, the process begins with hydrogen clouds collapsing to $10^{-3}\,M_\odot$ protostellar objects at the center of the halo. The protostar grows further by accretion, and hydrogen fusion turns on providing the stellar heat source for the stars. The accretion process stops once they grow to $\sim 100\,M_\odot$ due to the outward pressure of the hot ionizing photons leaving the stars. However, this standard picture could change if an additional heat source is present, such as dark matter annihilation. In this case, the collapse process could be halted due to dark matter heating, resulting in drastically different objects known as dark stars~\cite{Spolyar:2007qv, Freese:2008ur, Freese:2008hb}. 

Dark stars are composed almost entirely of ordinary matter (mostly hydrogen) and can be very luminous. The name ``dark star'' refers to their power source being dark matter annihilation rather than nuclear fusion for ordinary stars~\cite{Spolyar:2007qv, Freese:2008ur, Freese:2008hb, Iocco:2008xb, Iocco:2008rb, Freese:2008wh, Natarajan:2008db, Rindler-Daller:2014uja, Freese:2015mta, Rindler-Daller:2020yqe}. The products from dark matter annihilation can be trapped inside the hydrogen cloud where they heat up the baryonic gas, leading to the formation of a star in hydrostatic equilibrium. The dark matter power is spread out uniformly throughout the star that is too cool and diffuse to ignite nuclear fusion.

Previous studies~\cite{Spolyar:2007qv, Freese:2008ur, Freese:2008hb, Iocco:2008xb, Iocco:2008rb, Freese:2008wh, Natarajan:2008db, Rindler-Daller:2014uja, Freese:2015mta, Rindler-Daller:2020yqe} have focused on dark stars powered by weakly interacting massive particle (WIMP) dark matter. Dark stars could form for a wide range of WIMP masses and annihilation cross sections. These stars could grow extremely massive and luminous. Their radii are typically $10\,{\rm AU}$ and their surface temperatures $10^4\,{\rm K}$. They form with a mass of $1\,M_\odot$, but could continue to accrete surrounding matter and become as massive as $10^7\,M_\odot$ with luminosities as large as $10^{10}\,L_\odot$~\cite{Rindler-Daller:2014uja}, potentially making dark stars detectable with the James Webb Space Telescope~\cite{Freese:2010re}. Once the dark matter power runs out, these stars could collapse to black holes at high redshifts, thereby providing seeds for supermassive black holes in the early Universe~\cite{Freese:2010re,Ilie:2011da}.

In this work, we investigate the possibility of dark stars forming within the self-interacting dark matter (SIDM) scenario; see Ref.~\cite{Tulin:2017ara} for a review on SIDM. Dark matter self-interactions can transport heat and thermalize the inner halo~\cite{Spergel:1999mh,Dave:2000ar,Rocha:2012jg,Vogelsberger:2012ku,Kaplinghat:2015aga}. Taking into account the effects of baryons, SIDM predicts more diverse dark matter distributions in galaxies than collisionless dark matter such as WIMPs~\cite{Kaplinghat:2013xca,Elbert:2016dbb,Sameie:2018chj,Yang:2021kdf} -- it has been shown that SIDM can explain the full range of the observed diversity of rotation curves of spiral galaxies~\cite{Kamada:2016euw,Creasey:2017qxc,Ren:2018jpt}, a long standing challenge in astrophysics, see Refs.~\cite{Oman:2015xda,Salucci:2018eie,McGaugh:2020ppt} and references therein as well as Ref.~\cite{Salucci:2018hqu} for a recent review. The preferred self-interaction cross section per dark matter particle mass to address this {\it diversity problem} (as well as a host of other potential ``small-scale'' problems, see Refs.~\cite{DelPopolo:2016emo, Bullock:2017xww, Perivolaropoulos:2021jda} for reviews) is $\sigma_{\rm SI}/m_{\rm DM}\gtrsim1\,{\rm cm}^2/{\rm g}$ in galaxies, while the cross section is constrained to be $\sigma_{\rm SI}/m_{\rm DM} \lesssim 0.1\,{\rm cm}^2/{\rm g}$ in galaxy clusters~\cite{Kaplinghat:2015aga,Andrade:2020lqq}. Thus, astrophysical observations require a velocity dependent self-interaction cross section, which can be realized in many particle physics models~\cite{Tulin:2017ara}.

In order for a dark star to form, dark matter annihilation must start heating the hydrogen cloud faster than it cools due to the various baryonic cooling processes at some point during the collapse of the hydrogen cloud~\cite{Spolyar:2007qv}. When this happens, the dark matter power will slow down the contraction of the hydrogen cloud and a dark star is formed once the system has reached hydrostatic equilibrium. Two conditions must be met for dark matter heating to overcome the baryonic cooling processes: First, the heating rate from dark matter pair annihilation is proportional to the square of the dark matter density; in order for dark matter to be a sufficiently strong power source, its density in the collapsing hydrogen cloud must be sufficiently high. Second, the dark matter annihilation products must deposit their energy into the hydrogen cloud rather than escape the center of the halo or heat the remaining dark matter.

The purpose of this paper is to show that both of these necessary conditions for the formation of dark stars can be met in the SIDM scenario. Let us begin by discussing the first of these conditions. For collisionless dark matter such as WIMPs, the dark matter in the center of the halo follows a cuspy density profile, and the WIMP density increases further in the presence of a deepening baryonic potential as can be modeled by adiabatic contraction~\cite{Blumenthal:1985qy}. The resulting annihilation rate is high enough for heating from dark matter annihilation to overcome the cooling processes of the collapsing hydrogen cloud. The SIDM scenario must rely on a different mechanism to achieve a high enough dark matter density to power dark stars. SIDM-only simulations show that dark matter self-interactions generally lead to a shallow density core in the center of halos~\cite{Dave:2000ar,Rocha:2012jg}, in contrast to a density cusp in the WIMP case. In addition, the usual formalism of adiabatic contraction~\cite{Blumenthal:1985qy, 1984MNRAS.211..753B, Ryden:1987ska, 1980ApJ...242.1232Y, Gnedin:2004cx, Sellwood:2005pq} applies to collisionless dark matter only; SIDM, by definition, has efficient self-interactions.

A recent study~\cite{Feng:2020kxv} used a conducting fluid model~\cite{Balberg:2002ue} and showed that if baryons dominate the gravitational potential, their presence can speed up the onset of gravothermal collapse (``catastrophe") of an SIDM halo~\cite{Balberg:2002ue}, leading to a cuspy profile with a high central density~\cite{Ahn:2004xt, Koda:2011yb,Pollack:2014rja,Essig:2018pzq, Choquette:2018lvq,Nishikawa:2019lsc,Sameie:2019zfo}; see also Refs.~\cite{Kaplinghat:2013xca,Elbert:2016dbb,Sameie:2018chj,Feng:2020kxv,Yang:2021kdf}. Hydrodynamical simulations of the formation of protogalaxies that host the first stars have shown that baryons indeed dominate the central potential in these systems~\cite{Wise:2007bf}. Taking into account the baryonic influence on the gravothermal evolution of the SIDM halo, we will show that the central dark matter density in such protogalaxies could reach and exceed the values required for the formation of a dark star. 

Let us return to the second condition, i.e.~that the dark matter annihilation products must heat the baryonic gas. In the WIMP scenario, dark matter annihilates into standard model particles. In the dense protostellar baryon cloud, all electromagnetically-charged annihilation products thermalize quickly with the baryons and hence the power from dark matter annihilation is effectively transferred into the hydrogen gas. In typical SIDM model, dark matter annihilates almost exclusively into nonstandard-model particles. Nonetheless, efficient heating of the baryonic gas by dark matter annihilations is achieved if the primary annihilation products decay into electromagnetically-charged standard model particles before they leave the dense baryonic cloud where a dark star might form and before they interact with the remaining dark matter.

We will consider a simple but concrete SIDM model by introducing a fermionic dark matter candidate and a light scalar particle~\cite{Tulin:2013teo,Kahlhoefer:2017umn}. The scalar couples to both the dark matter candidate and the standard model fermions. While the dark matter particles will annihilate predominantly into the light scalars, this new scalar can have short enough decay lengths into standard model particles for SIDM annihilation to heat the protostellar hydrogen cloud. We will systematically study astrophysical constraints on this model and show that it can simultaneously achieve the observed dark matter density, the favored self-interacting cross section in galaxies, and evade direct and indirect detection constraints. The dark matter annihilation cross section in this model has interesting velocity dependence due to the interplay of $p$-wave suppression and Sommerfeld enhancement. This velocity dependence allows us to simultaneously achieve the observed relic density via canonical thermal freeze-out, large values of the annihilation cross section in the center of the $\sim 10^7\,M_\odot$ minihalos where dark stars could form, and suppressed annihilation cross sections at recombination allowing the model to evade stringent bounds from cosmic microwave background (CMB) observations.

The determination of the properties of dark stars in the SIDM scenario such as their luminosity, surface temperature, or size requires dedicated dynamical simulations of the gravitationally coupled evolution of the hydrogen and the SIDM distributions while simultaneously computing baryonic cooling rates and the effects of dark matter self-interactions and annihilation on both distributions. As a first step towards the study of the properties of dark stars in the SIDM scenario, we discuss the Lane-Emden equations describing a dark star once it has reached hydrostatic equilibrium. We solve these equations for the simplifying assumptions that the self-interactions in the dark star are fast enough to thermalize the SIDM such that the dark matter is described by an isothermal distribution. Without the aforementioned dedicated simulations, we do not know the dark matter density in the dark star; instead, we parametrize our solutions in terms of the SIDM-to-baryon mass ratio in the dark star. Our results suggest that the properties of dark stars in the SIDM scenario are similar to those in the WIMP scenario, although SIDM-powered dark stars might be slightly larger and cooler than WIMP-powered ones.

The remainder of the paper is organized as follows: In Sec.~\ref{sec:particle}, we introduce the SIDM model and study the viable parameter space. In Sec.~\ref{sec:core}, we investigate the possibility of achieving a sufficiently high dark matter density for powering a dark star via gravothermal evolution of the SIDM. In Sec.~\ref{sec:results}, we describe how to gain a first glimpse on the properties of SIDM powered dark stars under the assumptions described above and present numerical results. We summarize and conclude in Sec.~\ref{sec:con}. Throughout this work we use natural units with $\hbar = c = 1$. 

\section{Particle physics model of SIDM}
\label{sec:particle}

We consider an SIDM model extending the standard model by a gauge-singlet Dirac fermion $\chi$ which plays the role of the dark matter candidate and a gauge-singlet real scalar $\phi$ which mediates dark matter self-interactions as well as interactions between $\chi$ and standard model particles. In order to realize the velocity dependence of the dark matter self-interactions required to explain observations from dwarf galaxies to galaxy clusters~\cite{Kaplinghat:2015aga, Andrade:2020lqq}, i.e. strong dark matter self-interactions at galactic scales and weaker self-interactions at the galaxy cluster scale, we will consider a mediator much lighter than the dark matter candidate, i.e. $m_\phi \ll m_\chi$. If such a light mediator couples not only to the dark matter but also to standard model particles, strong constraints arise from direct and indirect detection experiments as well as from observations of the CMB~\cite{Bringmann:2016din, Cirelli:2016rnw, Choi:2016tkj, Kahlhoefer:2017umn, Chu:2018fzy, PandaX-II:2018xpz, Bernal:2019uqr, Ma:2019byo, Tsai:2020vpi, PandaX-II:2021lap}. As we will discuss in this section, the model we consider can simultaneously give rise to the observed dark matter relic density via thermal freeze-out, satisfy bounds from direct and indirect detection as well as the CMB, and, crucial for the possibility of a dark star, the energy from dark matter annihilation can be deposited in the baryonic gas in the environments where a dark star might form. The interplay of $p$-wave suppression and Sommerfeld enhancement due to the light mediator gives rise to an interesting velocity dependence of the dark matter annihilation cross section, which plays an important role in satisfying these requirements simultaneously.

We assume that the couplings of $\phi$ to dark matter particles $\chi$ and standard model fermions $f$ take the form
\begin{equation}
	\label{eq:couplings}
	\mathcal{L} \supset - y_\chi \overline{\chi} \chi \phi - y_f \sum_f \frac{\sqrt{2}m_f}{\langle V\rangle} i \overline{f} \gamma^5 f \phi \,,
\end{equation}
where $y_\chi$ ($y_f$) are dimensionless couplings between $\chi$ ($f$) and $\phi$, and we chose the couplings of $\phi$ to standard model fermions to be proportional to their standard model Yukawa couplings $\sqrt{2} m_f / \langle V\rangle$ with the mass of the fermion $m_f$ and the vacuum expectation value of the standard model Higgs boson $\langle V\rangle = 246\,$GeV. Throughout this work, we assume dark matter self-interactions to be much stronger than interactions of $\chi$ with the standard model, i.e.\ we take $y_f \ll y_\chi$. The coupling structure in Eq.~\eqref{eq:couplings} is maximally $CP$-violating: $\phi$ couples to $\chi$ like a $CP$-even scalar, but to standard model fermions like a $CP$-odd pseudo-scalar. Such a structure can, for example, be realized in a two Higgs doublet model extension of the standard model~\cite{Kahlhoefer:2017umn}. This coupling structure has various important phenomenological consequences. The dominant annihilation channel in this model is $\overline{\chi}\chi \to 2\phi$. For the $CP$-even $\overline{\chi}\chi\phi$ coupling, this process is $p$-wave suppressed such that the model can avoid constraints from indirect detection experiments and on the energy injection into the standard model bath in the early Universe while producing the observed dark matter relic density via thermal freeze-out. The $CP$-odd $\overline{f}\gamma^5f\phi$ coupling leads to velocity suppressed spin-independent direct detection cross sections~\cite{Fan:2010gt,Fitzpatrick:2012ix,Dent:2015zpa}, hence direct detection constraints are mitigated.

Let us discuss in some more detail the velocity dependence of the $\overline{\chi} \chi \to 2\phi$ annihilation cross section arising from the interplay of $p$-wave suppression and Sommerfeld enhancement. At tree level and in the nonrelativistic limit, the annihilation cross section is 
\begin{equation}
	\label{eq:DMtoMed}
	(\sigma v) = \frac{3\alpha^2_\chi v^2}{4 m^2_\chi} \sqrt{1-\frac{m^2_\phi}{m^2_\chi}} \,,
\end{equation}
where $\alpha_\chi \equiv y^2_\chi/4\pi$ and $v$ is the relative velocity of dark matter particles. In the limit $m_\phi \ll m_\chi$, the annihilation cross section receives nonperturbative corrections due to the effects of Sommerfeld enhancement,
\begin{equation}
	\label{eq:enhanced}
	\left(\sigma v \right)_{\rm eff} = S_p \times (\sigma v) \,,
\end{equation}
with
\begin{equation}
	\label{eq:Sommerfeld}
	\begin{split}
		S_p &= \frac{ \left(c-1\right)^2 + 4 \left(ac\right)^2}{1+4\left(ac\right)^2} S_s \,,\\
		S_s &= \frac{\pi \sinh(2\pi ac)}{a \left\{ \cosh(2\pi ac) - \cos\left[ 2\pi \sqrt{c-\left(ac\right)^2} \right] \right\} } \,,
	\end{split}
\end{equation}
where $a = v/2\alpha_\chi$ and $c = 6\alpha_\chi m_\chi/(\pi^2 m_\phi)$~\cite{Cassel:2009wt, Tulin:2013teo}. Thermal freeze-out (FO) occurs at $v \sim 0.1$, and the observed relic density is obtained for $( \sigma v )_{\rm eff}^{\rm FO} \simeq 6 \times 10^ {-26}\,{\rm cm}^3/$s. In the remainder of this work, we set the coupling $y_\chi$ to the value required to obtain the observed relic density via thermal freeze-out. 

\begin{figure}
    \includegraphics[width=\linewidth]{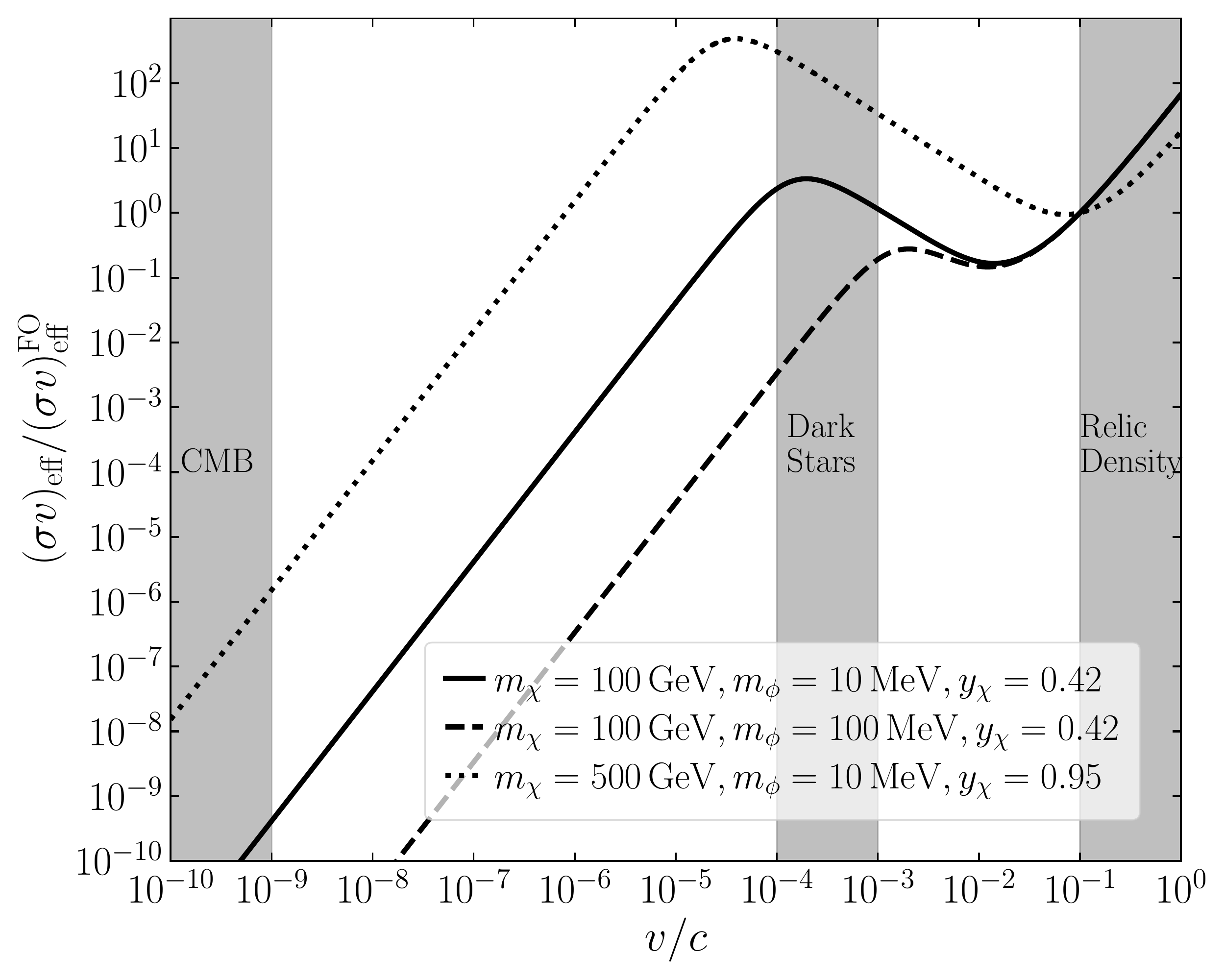}
    \caption{Velocity dependence of the annihilation cross section $(\sigma v)_{\rm eff}$, normalized to  $(\sigma v)^{\rm FO}_{\rm eff} = 6\times10^{-26}\,{\rm cm^3/s}$, for different choices of $m_\chi$ and $m_\phi$. The shaded areas show characteristic velocities during the recombination epoch relevant for CMB constraints, in dark stars, and at freeze-out relevant for setting the relic density.}
    \label{fig:sigv_vdependence}
\end{figure}

To illustrate the velocity dependence of $(\sigma v )_{\rm eff}$ arising from the interplay between $p$-wave suppression and Sommerfeld enhancement, we plot $(\sigma v)_{\rm eff} / (\sigma v )_{\rm eff}^{\rm FO}$ against the relative velocity $v$ for different choices of $m_\chi$ and $m_\phi$ in Fig.~\ref{fig:sigv_vdependence}. There are three overall regimes: For $v \gtrsim 0.1$, the Sommerfeld enhancement is negligible and the $p$-wave suppression leads to $( \sigma v )_{\rm eff} \propto v^2$. For velocities below $v \sim 0.1$, the enhancement becomes effective and the annihilation cross section scales as $( \sigma v )_{\rm eff} \propto 1/v$. However, for $v \lesssim m_\phi/m_\chi$, the Sommerfeld enhancement saturates and one recovers the usual $p$-wave scaling $(\sigma v)_{\rm eff} \propto v^2$. In Fig.~\ref{fig:sigv_vdependence}, we also indicate velocity regions relevant for different phenomenological aspects with the vertical gray bands. Thermal freeze out occurs at velocities $v \sim 0.1\textup{--}1$, as indicated by the band labeled ``relic density''; recall that throughout this work, we set the coupling $y_\chi$ to the value required to obtain the observed relic density. 

In environments where dark stars form, dark matter has velocities of order $v \sim 10^{-3}\textup{--}10^{-4}$ as indicated by the band labeled ``Dark Stars''. As we can see from Fig.~\ref{fig:sigv_vdependence}, due to the effects of Sommerfeld enhancement, the effective annihilation cross section at these temperatures can be {\it larger} than at freeze-out. If the Sommerfeld enhancement saturates at velocities $v \lesssim 10^{-3}$, which occurs for $m_\phi/m_\chi \lesssim 10^{-3}$, Sommerfeld enhancement is stronger than the $p$-wave suppression at the velocities relevant for the dark star, see the different cases shown in Fig.~\ref{fig:sigv_vdependence}. 

In addition, in the model we consider, dark matter annihilates into light scalars which ultimately decay into electromagnetically charged standard model particles. Observations of the CMB yield strong constraints on energy injection into the standard model bath at recombination times; for dark matter annihilation into electromagnetically charged standard model states this sets an upper limit of $\langle\sigma v \rangle_{\rm eff}^{\rm CMB}\lesssim 10^{-25}\,{\rm cm}^3/{\rm s} \times \left( m_\chi / 100\,{\rm GeV} \right)$~\cite{Planck:2018vyg,Slatyer:2015jla,Slatyer:2015kla}.\footnote{The future CMB-S4 observations~\cite{Abazajian:2019eic} will improve the sensitivity on the annihilation cross section by a factor of 3~\cite{Green:2018pmd}.}

The characteristic velocity of dark matter during the recombination epoch depends on the kinetic decoupling temperature $T_{\rm kd}$ of dark matter. For the model we consider, the elastic scattering processes $\chi f\rightarrow\chi f$ and $\chi\phi\rightarrow\chi\phi$ are present below the freeze-out temperature $T_{\rm FO} \sim m_\chi/20$. Elastic scattering processes of dark matter with standard model fermions are suppressed by the small standard model Yukawa couplings as well as our assumption of $y_f \ll y_\chi$, hence, $\chi\phi\rightarrow\chi\phi$ is the dominant process in setting $T_{\rm kd}$. In fact, the elastic scatterings between $\chi$ and $\phi$ particles keep dark matter in thermal equilibrium until the mediator becomes nonrelativistic~\cite{Feng:2010zp}, setting the scale for the kinetic decoupling of dark matter to $T_{\rm kd}\sim m_\phi$. The corresponding values for $T_{\rm kd}$ are generally slightly smaller than what is obtained in the WIMP scenario, and a later kinetic decoupling is a general feature of SIDM models (see e.g.\ Refs.~\cite{vandenAarssen:2012vpm, Cyr-Racine:2015ihg, Huo:2017vef, Huo:2019bjf, Egana-Ugrinovic:2021gnu}).

The temperature at which kinetic decoupling occurs has an impact on the dark matter velocity $v$ at recombination. As long as the dark matter is in thermal equilibrium, the dark matter velocity scales as $v \propto \sqrt{T}$ with the temperature of the standard model plasma $T$. However, after kinetic decoupling, i.e. for $T \lesssim T_{\rm kd}$, the dark matter velocity scales as $v \propto T$. Hence, the lower $T_{\rm kd}$, the higher the value for $v$ at recombination. We are mainly interested in mediator masses in the range $m_{\phi}\sim1\textup{--}100~{\rm MeV}$. For $m_\chi=100~{\rm GeV}$, this corresponds to dark matter velocities of $v \sim {10^{-9}}\textup{--}10^{-10}$ at recombination as indicated by the band labeled ``CMB'' in Fig.~\ref{fig:sigv_vdependence}. As long as the Sommerfeld enhancement of the annihilation cross section saturates at velocities $v \gg 10^{-9}$, corresponding to $m_\phi/m_\chi \gg 10^{-9}$, the effective dark matter annihilation cross section during recombination is much smaller than during freeze-out and the CMB bounds are satisfied.\footnote{The Sommerfeld enhancement features well-known resonances for fine-tuned combinations of the coupling $y_\chi$ and the ratio $m_\phi/m_\chi$~\cite{Arkani-Hamed:2008hhe}, see Eq.~\eqref{eq:Sommerfeld}. If the parameters of the models match the resonance condition, relevant bounds from the CMB can arise even for mass ratios as large as $m_\phi/m_\chi \sim 10^{-4}\textup{--}10^{-5}$. However, as these bounds arise only for extremely fine-tuned combinations of $y_\chi$ and $m_\phi/m_\chi$, we will neglect them in this work.} Another early Universe constraint on the model arises from energy injection during big bang nucleosynthesis (BBN), which we will discuss later in this section.

For the bounds on the coupling $y_f$ between $\phi$ and standard model fermions, the most stringent constraints stem from flavor physics measurements, in particular, constraints on charged kaon decays of the type $K^+ \rightarrow \pi^+ + \rm{invisible}$~\cite{Dolan:2014ska, Kahlhoefer:2017umn}, yielding
\begin{equation} 
    \label{eq:flavor}
    y_f \leq
    \begin{cases}
        8 \times 10^{-5}\,, & \, {\rm for}\,\, m_\phi \leq 100\,{\rm MeV}\,, \\
        2 \times 10^{-3}\,, & \, {\rm for}\,\, 100\,{\rm MeV} \leq m_\phi \leq 1\,{\rm GeV}\,.
    \end{cases}
\end{equation}

With the upper bounds on $y_f$ in Eq.~\eqref{eq:flavor}, we check if the energy from dark matter annihilation can be deposited in the baryonic gas in the environments where a dark star might form. As mentioned above, in our model dark matter annihilates predominantly into a pair of the new (light) scalars, $\overline{\chi}\chi \to 2\phi$. However, $\phi$ is not stable but decays into standard model particles -- if such decays happen before the $\phi$'s leave the dense hydrogen cloud and before the $\phi$'s interact with dark matter particles, dark matter annihilation can still heat the baryons and power a dark star. For $m_\phi < 2 m_e \approx 1\,$MeV, all decays of $\phi$ into pairs of electromagnetically charged standard model fermions are forbidden. The dominant decay mode then becomes the loop-suppressed $\phi \to \gamma\gamma$, with typical decay lengths much longer than the size of a dark star. For $m_\phi \gtrsim 1\,$MeV, on the other hand, decays into pairs of charged standard model states are allowed, and the associated decay length can be sufficiently small. We will be most interested in the mass range $1\,{\rm MeV} \lesssim m_\phi \lesssim 100\,$MeV, where the dominant decay mode will be into a pair of electrons.\footnote{For larger $m_\phi$, additional decay channels are relevant, e.g.\ into muons or pions, leading to even shorter decay lengths.} The associated decay width is  
\begin{equation}
    \label{eq:decay}
    \Gamma(\phi \rightarrow e^+e^-) = \frac{y_f^2}{4\pi} \frac{m_e^2}{\langle V\rangle^2} m_\phi \sqrt{1-\frac{4 m_e^2}{m_\phi^2}} \,,
\end{equation}
and the corresponding decay length is 
\begin{align}
    \label{eq:decaylength}
    \lambda_{\rm decay} &= \frac{\gamma}{\Gamma_{\phi \rightarrow e^+e^- }} \\
    &\sim 1\,{\rm\, AU} \left( \frac{8\times10^{-5}}{y_f} \right)^2 \left( \frac{10{\rm\,MeV}}{m_\phi} \right)^2 \left(\frac{m_\chi}{100{\rm \, GeV}} \right) \;, \nonumber
\end{align}
where $\gamma \sim m_\chi/m_{\phi}$ accounts for the relativistic boost of $\phi$'s from $\overline{\chi}\chi \to 2\phi$ annihilation, and $y_f = 8 \times 10^{-5}$ is the largest value allowed by the flavor physics constraints for $m_\phi \lesssim 100\,{\rm MeV}$, see Eq.~\eqref{eq:flavor}. The most important competing process is the scattering off dark matter particles, $\phi \chi \to \phi \chi$. If such scattering occurs before the $\phi$'s decays, the energy from dark matter annihilation will heat the dark matter rather than the baryons. The corresponding scattering cross section is $\sigma_{\phi\chi \to \phi\chi} \sim y_\chi^4 / 24\pi s$, where $s$ is the center-of-mass energy. The mean free path for $\phi$'s from $\overline{\chi}\chi \to 2\phi$ annihilation scattering off dark matter particles is
\begin{equation} \label{eq:scatteringlength}
    \lambda_{\rm scattering} \sim \frac{1}{n_\chi \sigma_{\phi\chi \to \phi\chi}} \sim 10^{11}\,{\rm AU} \left( \frac{10^{12}\,{\rm cm}^{-3}}{n_\chi} \right) \;,
\end{equation}
where $n_\chi$ is the dark matter number density with typical values of the order of $n_\chi \sim 10^{12}{\rm \,cm^{-3}}$ in a dark star as we will see below, we have set $m_\chi = 100\,$GeV, and fixed the coupling $y_\chi$ to yield the observed dark matter relic density. Comparing Eqs.~\eqref{eq:scatteringlength} and~\eqref{eq:decaylength} we see that $\phi$'s from dark matter annihilation decay into electrons much faster than they scatter off dark matter particles. For comparison to the decay length, a WIMP dark star is typically a few AU~\cite{Rindler-Daller:2014uja} in radius and, as we will discuss in Sec.~\ref{sec:results}, SIDM dark star could have similar or even larger sizes. Hence, requiring $\lambda_{\rm decay} < 1\,$AU, as we will do in the following, is a conservative criterion for the $\phi$'s resulting from dark matter annihilation to decay into electromagnetically charged standard model particles before they leave the baryon cloud or scatter off dark matter, ensuring that SIDM annihilation could power a dark star.    

\begin{figure}
    \includegraphics[width=\linewidth]{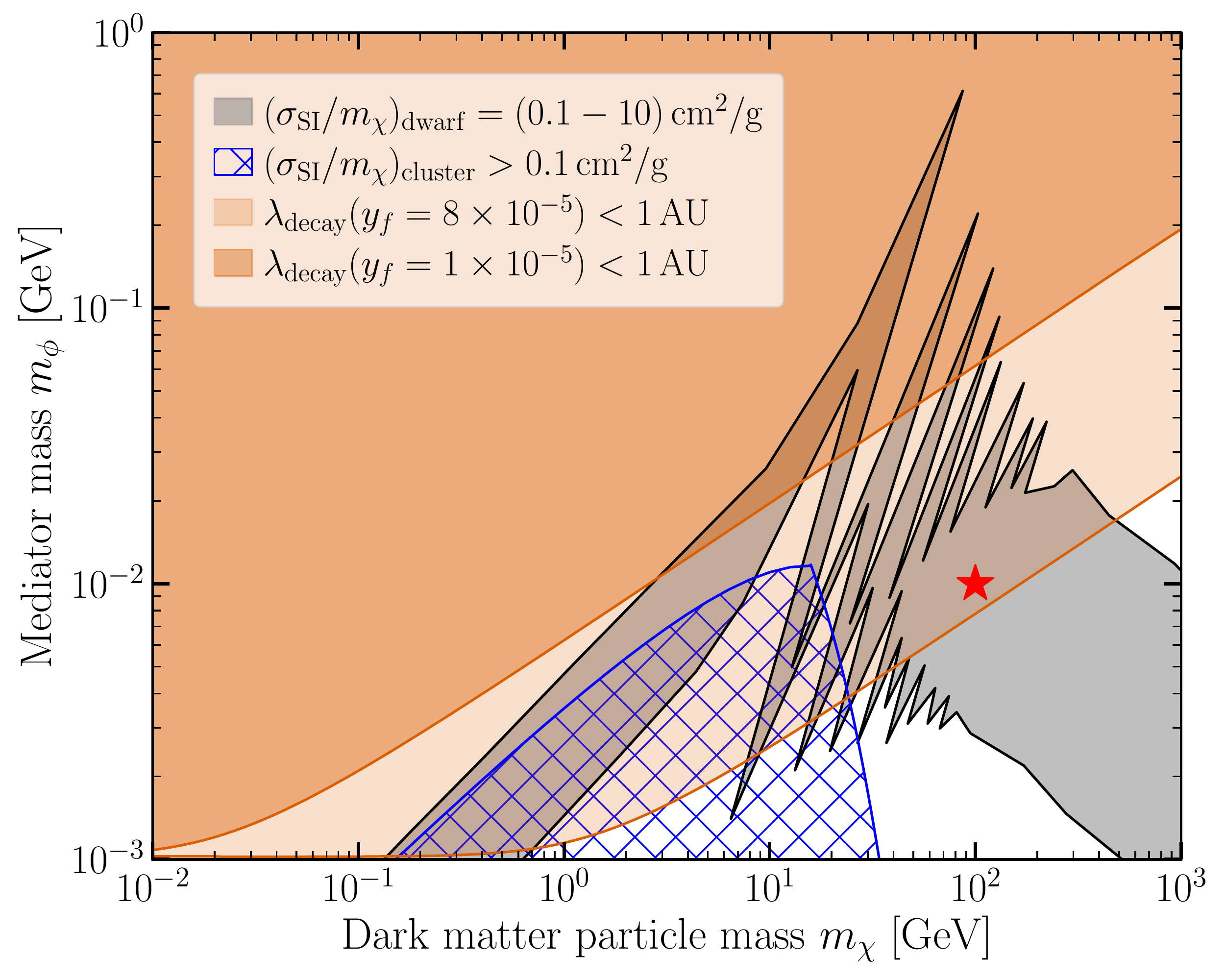}
    \caption{Different constraints on our model in the $m_\phi\textup{--}m_\chi$ plane as summarized in the final paragraph of Sec.~\ref{sec:particle}. In the gray-shaded band dark matter self-interactions at dwarf galaxy scales could alleviate the small-scale problems of collisionless dark matter. The blue hatched area is disfavored, in this region of parameter space dark matter self-interactions at galaxy-cluster scales lead to conflicts with observations. In the orange-shaded regions (the different orange shades are for different values of the coupling to standard model fermions as indicated in the legend), the decay length of the $\phi$'s from $\overline{\chi}\chi \to 2\phi$ annihilation are shorter than 1\,AU such that dark matter annihilation could power a dark star. The red star marks the benchmark point we will use in the remainder of this work.} 
    \label{fig:constraints}
\end{figure}

Another important constraint on the mediator lifetime arises from BBN. As the temperature becomes lower than the mediator mass, $\phi$ becomes nonrelativistic. The energy injection from $\phi$ decays could change the abundance of the light elements predicted in the standard scenario. We use Eq.~\ref{eq:decay} to estimate the $\phi$ lifetime as
\begin{equation} \label{eq:BBN}
        \tau_\phi\approx0.03{\rm\,s}\left(\frac{8\times10^{-5}}{y_f}\right)^2\left(\frac{10{\rm\,MeV}}{m_\phi}\right)\;.
\end{equation}
If the mediator decays before weak decoupling, $\tau_\phi \lesssim 1\,{\rm s}$, we are assured that the light element abundances will not be modified.\footnote{See Refs.~\cite{Hufnagel:2018bjp,Depta:2020zbh} for dedicated discussion of BBN constraints on models of this type, showing that $\tau_\phi\lesssim1\,{\rm s}$ is a conservative criterion.} As we can see from Eq.~\eqref{eq:BBN}, this condition is easily satisfied in our model.

Last but not least, we comment on indirect detection constraints. As discussed above, the dominant dark matter annihilation mode is $\chi \bar{\chi} \to 2\phi$. In the parameter region we are interested in for dark stars, the $\phi$'s decay into pairs of electrons with decay length $\lambda_{\rm decay} \lesssim 1\,$AU, such that for the purposes of indirect detection, the relevant annihilation mode is $\chi \bar{\chi} \to \phi\phi \to e^+ e^- e^+ e^-$. The strongest constraints on this annihilation channel in the dark matter mass range we are interested in arise from gamma rays produced in final state radiation of the electrons and positrons. As shown in Fig.~\ref{fig:sigv_vdependence}, the interplay of the $p$-wave suppression and the Sommerfeld enhancement can lead to rather large effective annihilation cross sections, $(\sigma v)_{\rm eff} \sim 10^{-25}\textup{--}10^{-24}\,{\rm cm}^3/$s, at the velocities relevant to indirect detection from the satellite galaxies of the Milky Way, $v \sim 10^{-4}$. Nonetheless, current indirect detection constraints do not exclude such cross section for $m_\chi \gtrsim 10\,$GeV, see e.g. Refs.~\cite{Essig:2009jx,Fermi-LAT:2015att,Hoof:2018hyn,Ando:2020yyk}, even before taking into account that these constraints assume cuspy dark matter profiles rather than the cored profiles one would expect for SIDM. At the galactic center of the Milky Way where $v\sim10^{-3}$, the electrons and positrons from the $\chi\bar{\chi}$ annihilation could scatter with starlight and produce gamma-ray signals; searches for such gamma-rays are sensitive to cross sections close to those giving rise to the observed relic density via freeze-out production~\cite{Kaplinghat:2015gha}. This constraint can be weakened if starlight and dark matter densities at the center are smaller than assumed in Ref.~\cite{Kaplinghat:2015gha}. The benchmark case with $m_\chi = 100\,$GeV and $m_\phi = 10\,$MeV we are going to use satisfies these constraints.

In Fig.~\ref{fig:constraints} we summarize the constraints on the parameter space discussed above. Recall that as a function of the dark matter mass $m_\chi$, we fix the coupling constant $y_\chi$ to obtain the observed dark matter relic abundance. The gray-shaded band (taken from Ref.~\cite{Kahlhoefer:2017umn}) shows the region of parameter space where the dark matter self-interaction cross section is in the range $\sigma_{\rm SI}/m_\chi = 0.1\textup{--}10{\rm\,cm^2/g}$ in dwarf halos (where $v \approx 5\times10^{-4}$), such that dark matter self-interactions can address the various small-scale problems of collisionless dark matter. In the blue hatched region, dark matter self-interaction at velocities corresponding to those in galaxy clusters are $\sigma_{\rm SI}/m_\chi \gtrsim 0.1\,{\rm cm^2/g}$; this region of parameter space is excluded by observations of galaxy clusters~\cite{Kaplinghat:2015aga,Andrade:2020lqq}. As discussed above, there are no relevant constraints in the region of parameter space shown in Fig.~\ref{fig:constraints} arising from direct detection or CMB constraints. In the orange-shaded regions in Fig.~\ref{fig:constraints}, the decay length of $\phi$'s from $\overline{\chi}\chi \to 2\phi$ annihilation is shorter than $1\,$AU, such that SIDM annihilation could power a dark star. The different shades of orange in Fig.~\ref{fig:constraints} correspond to different values of the coupling of $\phi$ to standard model fermions, note that both values of $y_f$ chosen in Fig.~\ref{fig:constraints} are allowed by the flavor physics constraints discussed above. Considering all of these constraints, Fig.~\ref{fig:constraints} demonstrates that there is a viable region of parameter space where our model exhibits the desired dark matter phenomenology, experimental and observational bounds are satisfied, and the energy from dark matter annihilation can be deposited in the baryonic gas in the environments where a dark star might form. In the remainder of this work, we use a benchmark case from this preferred region: $m_\chi = 100\,$GeV, $m_{\phi} = 10\,$MeV, and $y_f = 8\times10^{-5}$, as highlighted by the red star in Fig.~\ref{fig:constraints}. 

\section{Gravothermal evolution of SIDM halos}
\label{sec:core}

As discussed in the Introduction, there are two necessary conditions for the formation of dark stars. In the previous section we showed that the first condition can be met in a consistent SIDM model, namely that the power from dark matter annihilation can be deposited in the hydrogen cloud. In this section, we argue that the second condition can be satisfied, i.e.\ that the dark matter density in the collapsing hydrogen cloud can be high enough for dark matter heating to overcome the baryonic cooling processes. 

The heating rate resulting from dark matter annihilation integrated within a sphere of radius $R$ is
\begin{equation}
    \label{eq:heatingrate}
    L_{\rm DM} = \frac{\left(\sigma v\right)_{\rm eff}}{2 m_\chi}\int^R_0\mathrm{d}{r}\,4\pi r^2\rho_{\rm DM}^2({r})\,,
\end{equation}
where $\rho_{\rm DM}$ the dark matter density and $\left(\sigma v\right)_{\rm eff}$ is the annihilation cross section which, for the SIDM model, is expressed by Eqs.~\eqref{eq:DMtoMed}--\eqref{eq:enhanced}. Up to $\mathcal{O}(1)$ factors\footnote{In the case of WIMPs, typically a third of the heating rate will be lost to neutrinos escaping the star. Furthermore, for Majorana fermion dark matter an additional factor 2 must be included compared to Dirac fermions.}, Eq.~\eqref{eq:heatingrate} is also applicable for a dark star powered by WIMP annihilation~\cite{Spolyar:2007qv}. Hence, we can use the results from the studies of dark stars in the WIMP scenario to estimate the required dark matter densities for dark matter heating to overcome the baryonic cooling processes. As discussed in Sec.~\ref{sec:particle}, at the dark matter velocities relevant in a dark star, $\left(\sigma v\right)_{\rm eff}$ in the SIDM scenario can be much larger than the benchmark value $(\sigma v)_{\rm eff} \sim 3 \times 10^{-26}\,{\rm cm}^3/$s appropriate for Majorana fermion WIMPs produced via $s$-wave freeze-out. Thus, requiring SIDM to reach larger densities than the minimal dark matter densities previous studies for WIMP powered dark stars found for dark star formation is a {\it conservative} criterion -- due to the larger $(\sigma v)_{\rm eff}$, SIDM is more effective at heating the baryons than WIMPs for the same $\rho_{\rm DM}$.

In the WIMP case, the dark matter density in the collapsing hydrogen cloud is controlled by adiabatic contraction. This leads to a relation between $\rho_{\rm DM}$ and the gas number density $n_{\rm gas}$ at the edge of the baryonic core as $\rho_{\rm DM} \approx 5\,{\rm GeV}\,{\rm cm}^{-3} \left( n_{\rm gas} / {\rm cm}^{-3} \right)^{0.81}$~\cite{Spolyar:2007qv}. For a $m_\chi = 100\,$GeV WIMP mass and $(\sigma v)_{\rm eff} = 3 \times 10^{-26}\,{\rm cm}^3/$s, Ref.~\cite{Spolyar:2007qv} found that dark matter heating overcomes the baryonic cooling processes for a gas density of $n_{\rm gas} \sim 10^{13}\,{\rm cm}^{-3}$~\cite{Spolyar:2007qv}. The baryonic core has a size of $r \sim 4 \times 10^{-4}\,$pc at that stage, and the dark matter density at the edge of the baryonic core is $\rho_{\rm DM} \sim 2 \times 10^{11}\,{\rm GeV/cm}^3$.

Can such densities be reached in an SIDM halo? Consider a $\sim10^{7}\,M_\odot$ halo at redshift $z\sim20$, which simulations suggest to be the birth place of the first stars in the Universe~\cite{Abel:2001pr,Gao:2006ug,Wise:2007bf}. For these halos, the scale radius and characteristic density are typically $r_s \sim 50{\rm\,pc}$ and $\rho_s\sim100{\rm\,GeV/cm^3}$, respectively. For a pure SIDM halo, self-interactions produce a shallow density core with the central density approaching $\rho_s$, which is many orders of magnitude lower than the threshold densities required to power a dark star.

However, the presence of the baryonic gas in the protogalaxies that host the first stars dramatically changes the gravothermal evolution of the halo, resulting in high central dark matter densities. Reference~\cite{Feng:2020kxv} performed simulations of the evolution of an SIDM halo under the influence of the baryonic potential in such protogalaxies. The initial conditions in Ref.~\cite{Feng:2020kxv} were based on the result of Ref.~\cite{Wise:2007bf} for such protogalaxies: the initial dark matter profile for a $3.5\times10^{7}\,M_\odot$ halo at $z\approx16$ is fit by a Navarro-Frenk-White (NFW) profile~\cite{Navarro:1995iw} with scale radius $r_s\approx73{\rm\,pc}$ and characteristic density $\rho_s \approx 98.9{\rm\,GeV/cm^3}$, while the gas density profile is fit by a power law, $\rho_{\rm gas} \propto r^{-2.4}$.

\begin{figure}
    \includegraphics[width=0.9\linewidth]{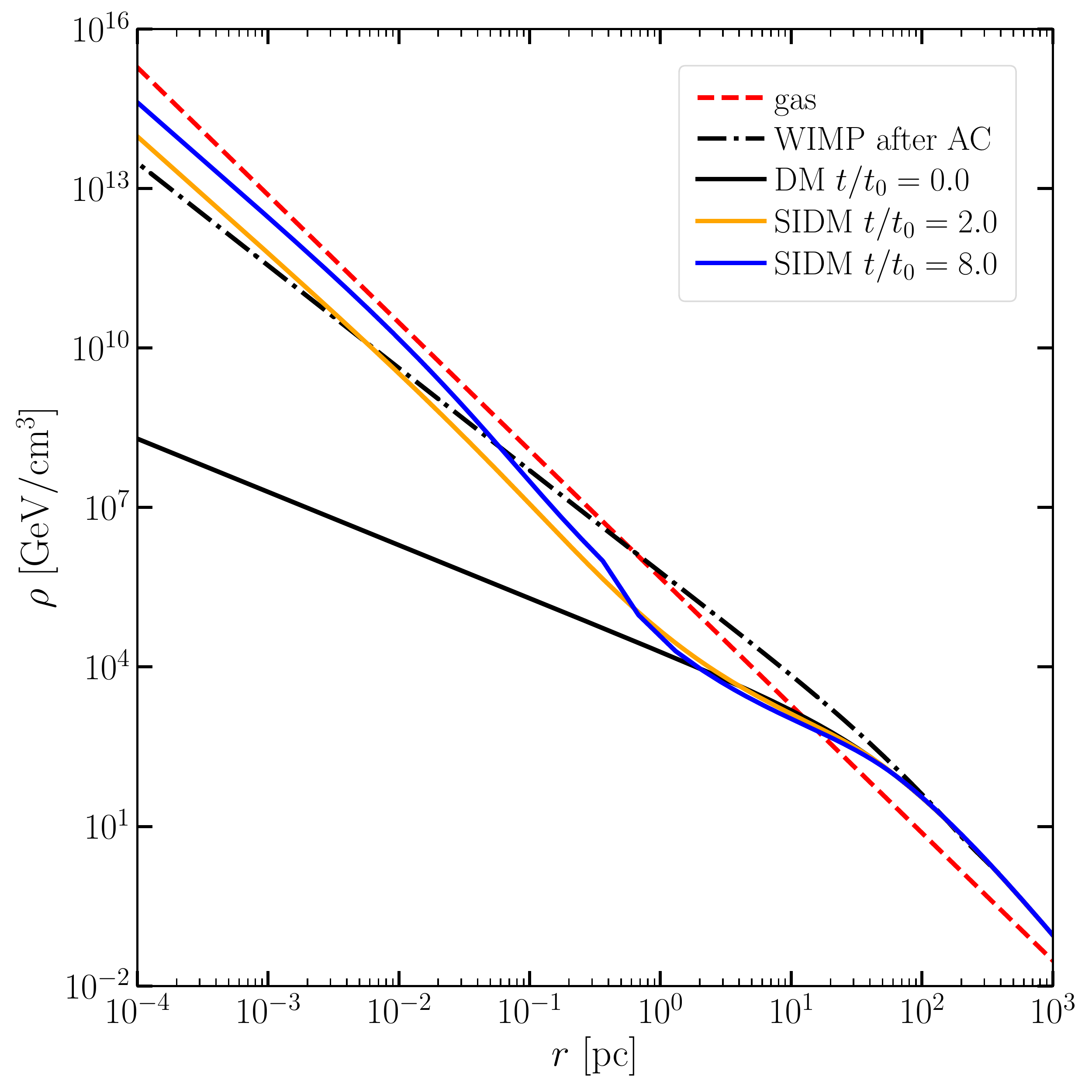}
    \caption{Gravothermal evolution of an SIDM halo in the presence of a static gravitational potential source by baryonic gas (measured in units of the characteristic time $t_0 = 1/\sqrt{4\pi G \rho_s} \approx 2.6{\rm\,Myr}$). The different lines show the dark matter density profiles at $t/t_0=0$ (solid black), $2$ (solid orange) and $8$ (solid blue); based on SIDM simulations in Ref.~\cite{Feng:2020kxv}, where the gas (dashed red) and initial dark matter (solid black) density profiles are from hydrodynamical cosmological simulations of protogalaxies~\cite{Wise:2007bf}. For comparison, the dash-dotted black line shows the density profile for a WIMP halo with the same initial dark matter profile undergoing adiabatic contraction by the same gas profile.}
    \label{fig:collapse}
\end{figure}

Figure~\ref{fig:collapse} shows the evolution of SIDM density profiles in the presence of the baryonic potential, based on the simulations in Ref.~\cite{Feng:2020kxv}, where time is measured in units of the characteristic time $t_0 = 1/\sqrt{4\pi G \rho_s} \approx 2.6{\rm\,Myr}$, a convenient timescale to measure the evolution of self-gravitating systems. As we can see in Fig.~\ref{fig:collapse}, the dark matter density in the inner regions of the halo increases rapidly, for example, at $t=2t_0$, the density at a radius $r \sim 4 \times 10^{-4}\,$pc reaches $\rho_{\rm DM} \gtrsim 10^{12}\,{\rm GeV/cm}^3$, {\it larger} than the threshold density for a WIMP dark star discussed above. For a more global comparison of the dark matter densities, we also show the density profile for WIMP dark matter after adiabatic contraction starting from the same NFW profile and under the effect of the same gas density as for the SIDM case in Fig.~\ref{fig:collapse}. This result is obtained using the publicly available \texttt{contra} code~\cite{Gnedin:2004cx}. The density of SIDM at $t=2t_0$ already surpasses the density of WIMPs after adiabatic contraction in the inner region ($r \lesssim 10^{-2}\,$pc) of this halo. Thus, via the gravothermal evolution of SIDM under the influence of the gravitational potential of the baryons, the inner region of the SIDM halo could provide sufficiently high dark matter densities to form a dark star and the influence of the gravitational potential of the baryons dramatically changes the gravothermal evolution of the SIDM halo. An SIDM-only halo would have first formed a shallow isothermal core, and the gravothermal core collapse (or gravothermal catastrophe) would only have occurred on much longer timescales, $t \sim (10^2\textup{--}10^3) t_0$~\cite{Balberg:2002ue, Ahn:2004xt, Essig:2018pzq}; see also Refs.~\cite{Outmezguine:2022bhq,Yang:2022zkd,Yang:2022hkm} for recent work on the gravothermal evolution of SIDM-only halos. The deep baryonic potential can trigger the onset of gravothermal collapse of the SIDM halo~\cite{Kaplinghat:2013xca, Elbert:2016dbb, Sameie:2018chj, Feng:2020kxv, Yang:2021kdf} on timescale of order $t_0$, leading to extremely high central dark matter densities~\cite{Balberg:2002ue, Ahn:2004xt, Koda:2011yb,Pollack:2014rja,Essig:2018pzq, Choquette:2018lvq,Nishikawa:2019lsc, Huo:2019yhk, Turner:2020vlf, Correa:2020qam, Sameie:2021ang,Zeng:2021ldo, Silverman:2022bhs}.

The results in Fig.~\ref{fig:collapse} stem from the simulations in Ref.~\cite{Feng:2020kxv}. These simulations use a semianalytical conducting fluid model and assume a static baryonic potential, allowing Ref.~\cite{Feng:2020kxv} to achieve sufficiently high spatial resolution to resolve the central halo, necessary for us to evaluate the conditions of dark star formation. The overall results are consistent with the findings in recent hydrodynamical N-body simulations~\cite{Sameie:2021ang}, showing that in the presence of a highly concentrated baryon profile, the inner SIDM halo is more dense than its collisionless dark matter counterpart. 

In addition, Ref.~\cite{Feng:2020kxv} considers a fixed dimensionless cross section parameter $(\sigma_{\rm SI}/m_\chi)(r_s\rho_s)=0.2$. Taking the halo parameters $\rho_s=73{\rm\,pc}$ and $\rho_s=98.9{\rm\,GeV/cm^3}$, this corresponds to $\sigma_{\rm SI}/m_\chi \approx 7.3\,{\rm cm}^2/$g, which is within the favorable range of $\sigma_{\rm SI}/m_\chi$ shown in Fig.~\ref{fig:constraints}. Given the rapid growth rate of the central SIDM density indicated in Fig.~\ref{fig:collapse}, we expect that smaller values of $\sigma_{\rm SI}/m_\chi$ would still lead to sufficiently high dark matter densities for dark matter heating to overcome baryonic cooling processes. However, determining the range of self-interaction strengths for which the evolution of the SIDM halo under the influence of the collapsing hydrogen cloud leads to sufficiently high dark matter densities for dark stars would require running a suite of simulations which we leave for future work.

\section{Properties of SIDM dark stars}
\label{sec:results}

In Secs.~\ref{sec:particle} and~\ref{sec:core} we have shown that the necessary conditions for the formation of dark stars can be met in the SIDM scenario, namely that the dark matter density in the environments where dark stars would form can be high enough to make dark matter annihilation a sufficiently strong power source to overcome the baryonic cooling processes, and that the energy from dark matter annihilation can be deposited in the hydrogen cloud to power a dark star. Once the energy injection from dark matter annihilation surpasses the energy loss due to baryonic cooling, the dissipative collapse of the gas slows down. Eventually, the baryons can reach hydrostatic equilibrium, resulting in a dark star.

As discussed in Sec.~\ref{sec:core}, the evolution of the baryonic and SIDM densities are deeply intertwined, hence, determining the properties of SIDM dark stars such as their luminosity, surface temperature, or size requires detailed simulations of the coupled evolution of the baryonic and SIDM fluids including the various baryonic cooling mechanisms, the heat transport in the SIDM component, and the coupled gravothermal evolution of the baryonic and SIDM fluids. Such dedicated numerical studies are beyond the scope of this work. In order to gain a first glimpse on the properties of SIDM powered dark stars and how they compare with their counterparts in the WIMP scenario, we will make the simplified assumption that the SIDM distribution in the dark star is isothermal. This assumption is motivated by the fact that while SIDM efficiently self-scatters (and hence self-thermalizes), the interactions of SIDM with baryons are much weaker. Furthermore, we initially treat the SIDM-to-baryon mass ratio inside the photosphere of the dark star $f_{\rm DM}$ as a free parameter, which we later fix when discussing numerical results using a prescription described below.

The dark matter and baryonic gas distributions in the dark star are described by 
\begin{equation}
    \label{eq:static}
    \nabla p_{\rm DM}= - \rho_{\rm DM}\nabla \Phi\,, \quad \nabla p_{\rm gas}= - \rho_{\rm gas}\nabla \Phi\,,
\end{equation}
where $\rho_{\rm DM}$ ($\rho_{\rm gas}$) and $p_{\rm DM}$ ($p_{\rm gas}$) are dark matter (gas) density and pressure, respectively, and $\Phi$ is the gravitational potential. The two components are coupled via the Poisson equation,
\begin{equation}
    \label{eq:poisson}
    \nabla^2 \Phi = 4\pi G (\rho_{\rm gas}+\rho_{\rm DM})\,.
\end{equation}
We assume that the baryonic gas satisfies a polytropic equation of state, $p_{\rm gas} = K_{\rm gas}\,\rho_{\rm gas}^{1+1/n}$, where $n$ is the polytropic index and $K_{\rm gas} \equiv p_{{\rm gas}}(0)/\rho_{{\rm gas}}^{1+1/n}(0)$ is a constant of proportionality evaluated at radius $r = 0$. Furthermore, we assume that the SIDM profile is described by an isothermal distribution, corresponding to a polytrope with $n = \infty$ or $p_{\rm DM} = K_{\rm DM} \rho_{\rm DM}$. Since dark stars are dominantly composed of baryons ($f_{\rm DM} \ll 1$), the baryonic mass dominates the gravitational potential with $\Phi(R) \approx - G M_\star/R$, where $M_\star$ is the baryonic mass enclosed in a given stellar radius $R$. Assuming that the dark matter in the dark star is virialized, we have $K_{\rm DM}=GM_{\star}/(3R)$. In addition, we evaluate the SIDM annihilation cross section in the star using the dark  matter root-mean-square speed $\sqrt{\langle v^2\rangle }= \sqrt{GM_\star/R}$.

\begin{figure}
    \includegraphics[scale=0.3]{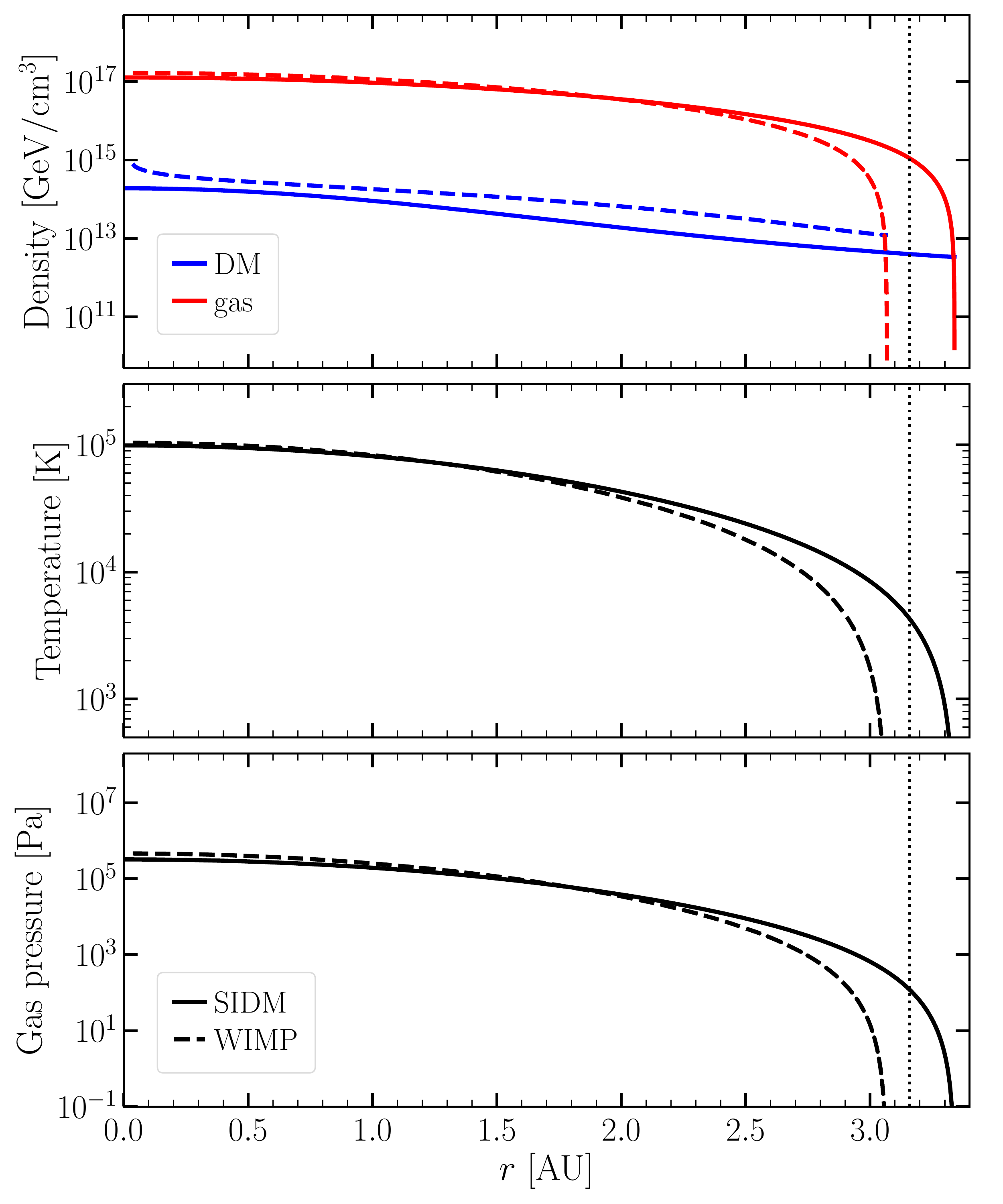}
    \caption{The profiles of gas (red) and dark matter (blue) densities (top), temperature (middle) and pressure (lower) of $10\,M_\odot$ SIDM-powered (solid lines) and WIMP-powered (dashed lines) dark stars. Both cases are for a dark matter particle mass of $m_\chi = 100\,$GeV. For the SIDM-powered dark star, we set the SIDM-to-baryon mass ratio in the dark star to $f_{\rm DM} = 7.4\times10^{-4}$ such that both dark stars have the same total luminosity, $L_{\rm eff} = 1.5\times10^5\,L_\odot$. The vertical dotted black line indicates the photosphere radius of the SIDM-powered dark star, $R_\star = 3.2{\rm\,AU}$, and the photosphere temperature of the SIDM dark star is $T_{\rm eff}=4300\,$K. For the WIMP dark star, we find $R_\star = 2.9\,$AU and $T_{\rm eff} = 4600\,$K.}
    \label{fig:profiles}
\end{figure}

Assuming spherical symmetry, we obtain dark matter and stellar density profiles iteratively in the following way. We take $M_{\star}$ and $f_{\rm DM}$ as input parameters and make an initial guess for the stellar radius $R$, then solve Eqs.~\eqref{eq:static}-\eqref{eq:poisson}, supplemented with the equations of state for SIDM and gas. Following the prescription in~\cite{2007nmai.conf.....B}, we interpolate the polytropic index of the gas profile between $n=3/2$ (fully convective star) and $n=3$ (fully radiative star), depending on the gas density. We write the pressure profile inside the star as a sum of the matter and the radiation pressures,
\begin{equation}
    \label{eq:pressure}
    p_{\rm gas}(r) = \frac{\rho_{\rm gas}(r) }{\mu\,m_u}k_B T_{\rm gas}(r) + \frac{1}{3}a T_{\rm gas}^4(r)\,,
\end{equation}
where $\mu\approx0.588$ is the mean molar mass of gas particles~\cite{Freese:2008wh}, $m_u \approx 0.933\,$GeV is the atomic mass unit, $k_B$ is Boltzmann's constant, and the radiation constant $a=4\sigma_{\rm SB}$ is related to the Stefan-Boltzmann constant $\sigma_{\rm SB}$ (we remind that we use units with $c = 1$). We invert Eq.~\eqref{eq:pressure} to obtain the temperature profile $T_{\rm gas}(r)$. From $T_{\rm gas}(r)$ and $\rho_{\rm gas}(r)$ profiles, we calculate the Rosseland mean opacity $\kappa$ by interpolating zero-metallicity tables~\cite{1996ApJ...464..943I, 1991ApJS...76..759L}. The photosphere radius $R_\star$ is given by the condition
\begin{equation}
    \label{eq:photo}
    \kappa p_{\rm gas}(R_\star) = \frac{2}{3} \frac{{\rm d}\Phi(r)}{{\rm d}r}\bigg\vert_{r=R_\star}\,.
\end{equation}
At the photosphere radius $R_\star$, we calculate the effective temperature as $T_{\rm eff}=T_{\rm gas}(R_\star)$ and the effective luminosity,
\begin{equation}
    \label{eq:leff}
    L_{\rm eff} = 4\pi \sigma_{\rm SB} R_\star^2\,T_{\rm eff}^4\,.
\end{equation}
We compare $L_{\rm eff}$ with the total luminosity $L_{\rm DM}$ produced from dark matter annihilation, see Eq.~\eqref{eq:heatingrate}. We iteratively adjust the stellar radius $R$ and repeat the procedure described above until $L_{\rm DM}$ and $L_{\rm eff}$ match: $|L_{\rm DM}-L_{\rm eff}|/(L_{\rm DM}+L_{\rm eff})<0.05$. We do not consider nuclear fusion as a power source; the radii of the smallest dark stars considered in the iterative process are much larger than those of ordinary Pop-III stars and the cores of dark stars are not dense and hot enough to start nuclear fusion.

In Fig.~\ref{fig:profiles} we show the gas and dark matter density profiles (top) as well as the temperature (middle) and pressure (bottom) profiles for $M_\star = 10\,M_\odot$ SIDM-powered (solid) and WIMP-powered (dashed) dark stars. The solution for the WIMP-powered dark star is found following the procedure outlined in Ref.~\cite{Spolyar:2007qv}; in particular, the dark matter density profile in the dark star is obtained by adiabatic contraction of a $10^7\,M_\odot$ NFW halo at redshift $z \sim 20$ with concentration parameter $c_{\rm NFW}=3$. We assume a WIMP of mass $m_\chi = 100\,$GeV and set the annihilation cross section to $(\sigma v)_{\rm eff} = 3 \times 10^{-26}\,{\rm cm}^3/$s, yielding the observed dark matter relic density for a Majorana fermion WIMP annihilating via $s$-wave processes. For the SIDM dark star, we take the benchmark case from the end of Sec.~\ref{sec:particle}: $m_\chi = 100\,$GeV dark matter particle mass, $m_\phi = 10\,$MeV mediator mass, and the annihilation cross section in the dark star is calculated using Eq.~\eqref{eq:enhanced} fixing the effective annihilation cross section at freeze-out to $(\sigma v)_{\rm eff} = 6 \times 10^{-26}\,{\rm cm}^3/$s such that one obtains the observed dark matter relic density for the Dirac fermion dark matter candidate $\chi$. In order to match the total luminosity of the SIDM-powered dark star to that of the WIMP-powered dark star, $L_{\rm tot} = 1.5\times 10^5\,L_\odot$, we set the SIDM-to-baryon mass ratio to $f_{\rm DM} = 7.4 \times 10^{-4}$.

In Fig.~\ref{fig:profiles}, we also indicate the photosphere radius $R_\star = 3.2\,$AU of the SIDM dark star with the vertical dotted black line. The dark matter density drops gradually with increasing radius $r$, and it extends beyond the photosphere. For the gas, its density, temperature, and pressure profiles follow a similar trend within $R_\star$, but drop quickly for $r \gtrsim R_\star$. Comparing the profiles of the SIDM- and WIMP-powered dark stars, we can note that their properties are rather similar -- while the SIDM dark star has $R_\star = 3.2\,$AU, the WIMP dark star is slightly smaller/denser with $R_\star = 2.9\,$AU. In turn, the SIDM dark star is somewhat colder; its effective photosphere temperature is $T_{\rm eff} = 4300\,$K, while the photosphere temperature of the WIMP dark star is $T_{\rm eff} = 4600\,$K.

\section{Summary and Conclusions}
\label{sec:con}

We have shown that the necessary conditions for the formation of dark stars can be met in the SIDM scenario. In particular, this work demonstrates that in a concrete SIDM particle physics model satisfying all current constraints, the power from dark matter annihilation can be deposited in the baryonic gas. When the baryonic gas cloud in the $(10^6\textup{--}10^8)\,M_\odot$ protohalos thought to be the birthplaces of the first star collapses, the deepening gravitational potential can lead to gravothermal evolution of the SIDM fluid resulting in very high central dark matter densities. These densities are sufficiently high for the power from dark matter annihilation deposited in the baryons to overcome all baryonic cooling processes, giving rise to the appropriate conditions for forming a dark star. While investigating the detailed properties of such dark stars in the SIDM scenario once they have reached hydrostatic equilibrium is beyond the scope of this work, we have given a first estimate of their properties assuming that the SIDM distribution in the dark star is isothermal and using the SIDM-to-baryon mass ratio in the dark star as an input parameter.

There are several relevant topics that require further investigations. In order to form a dark star, a high central density in the dark matter halo is required. In SIDM models, the presence of the baryonic gas can deepen the gravitational potential and increase the density accordingly. This effect has been confirmed in hydrodynamical simulations for Milky Way-like systems at low redshifts~\cite{Sameie:2021ang}. In order to study the properties of dark stars in the SIDM scenario such as their luminosity and surface temperature, dedicated dynamical simulations are needed. Ideally, such simulations would model the evolution of the hydrogen and SIDM distributions starting from initial conditions as appropriate for $(10^6\textup{--}10^8)\,M_\odot$ protohalos at redshifts $z \sim 20$ before the hydrogen cloud in these halos begins to contract. The simulations would need to include modeling of the relevant baryonic cooling processes and chemistry, the dark matter self-interactions, heating from dark matter annihilations, and the gravitational interactions to track the coupled gravothermal evolution of both the baryons and the dark matter. Similar simulations have been undertaken for studying the early stages of dark star formation in the WIMP dark matter scenario, see, e.g., Refs.~\cite{Stacy:2011rp, Smith:2012ng, Stacy:2013xwa}. We stress that in the SIDM case, dark matter cannot be modeled with the collisionless N-body approach that is applied for WIMP dark matter, as these simulations must account for the effects of dark matter self-interactions. It is also very challenging to resolve the central few-AU region where a dark star would form even for ($10^6\textup{--}10^8)\,M_\odot$ dark matter minihalos at redshifts $z \sim 20$: for example, the $3.5 \times 10^7\,M_\odot$ halo at $z \approx 16$ we considered in Sec.~\ref{sec:core} has a (NFW) scale radius of $r_s = 73\,{\rm pc} = 1.5 \times 10^7\,$AU. A possible solution is to use a series of simulations to model the formation of dark stars in the SIDM scenario. For instance, the detailed properties of dark stars such as their luminosity and surface temperature could be modeled with dedicated stellar evolution codes, for which the only necessary information regarding the dark matter physics is the heating rate which can be computed once the evolution of the dark matter density and temperature is understood, see, e.g., Refs.~\cite{Rindler-Daller:2014uja,Freese:2015mta} for such an approach in the WIMP dark matter scenario.

Further interesting questions that warrant future work include the possible growth of dark stars via baryon accretion~\cite{Spolyar:2007qv, Freese:2010re} or dark matter capture~\cite{Freese:2008ur, Iocco:2008xb}. In SIDM, dark matter particles could experience self-capture due to the large self-interaction cross section~\cite{Zentner:2009is, Kouvaris:2011gb}. Such a self-capture process could affect the formation and evolution of dark stars. It would also be interesting to study observational prospects of SIDM dark stars with, for example, the James Webb Space Telescope~\cite{Gardner:2006ky}.

\begin{acknowledgments}
We thank Yi-Ming~Zhong for proving simulation data used in Fig.~\ref{fig:collapse}, and Oleg~Gnedin, Manoj~Kaplinghat, Ben~Sheff, and Monica~Valluri for helpful discussions. 
We acknowledge the kind hospitality of the Leinweber Centre for Theoretical Physics (LCTP) at the University of Michigan, where this work was initiated. 
Y.W.~is supported by the MICDE catalyst grant at University of Michigan and the DoE grant No.~DE-SC007859. 
S.B.~is grateful for support via the Clark~Fellowship at the Stanford Institute for Theoretical Physics, NSF Grant No.~PHYS-2014215, DoE HEP QuantISED award No.~100495, and the Gordon and Betty Moore Foundation Grant No.~GBMF7946. 
L.V.~acknowledges support from the NWO Physics Vrij Programme ``The Hidden Universe of Weakly Interacting Particles'' with project number 680.92.18.03 (NWO Vrije Programma), which is (partly) financed by the Dutch Research Council (NWO), as well as support from the European Union's Horizon 2020 research and innovation programme under the Marie Sk{\l}odowska-Curie grant agreement No.~754496 (H2020-MSCA-COFUND-2016 FELLINI). 
K.F.~is grateful for support via her Jeff \& Gail Kodosky Endowed Chair in the Department of Physics at the University of Texas, Austin. K.F.~acknowledges support from the Vetenskapsr\r{a}det (Swedish Research Council) through contract No.~638-2013-8993, DoE Grant DE-SC-0022021 at the University of Texas, Austin, as well as prior funding from DoE grant No.~DE-SC007859 at the University of Michigan. 
H.B.Y.~acknowledges support from DoE under grant No.~DE-SC0008541 and the John Templeton Foundation (ID\# 61884). The opinions expressed in this publication are those of the authors and do not necessarily reflect the views of the John Templeton Foundation.
\end{acknowledgments}

\bibliography{SIDM}

\end{document}